\documentclass[nofootinbib,floatfix,superscriptaddress]{revtex4}        
\usepackage{graphicx}
\usepackage{epsfig}
\usepackage{bm}
\usepackage[T1]{fontenc}
\usepackage[latin9]{inputenc}
\usepackage{amssymb}
\usepackage{float}
\usepackage{subfig}
\usepackage{amsmath}
\usepackage{dcolumn}
\usepackage{cancel}
\usepackage[colorlinks]{hyperref}
\usepackage[usenames,dvipsnames]{color}
\hypersetup{
     breaklinks=true,
    pdfstartview={FitH},    
    colorlinks=true,       
    linkcolor=blue,          
    citecolor=red,        
    filecolor=magenta,      
    urlcolor=blue,           
    anchorcolor=green,      
    linktocpage=true
}

\newcommand{\be}{\begin{equation}}
\newcommand{\ee}{\end{equation}}

\begin{document}

\title{Constraining Barrow entropy-based Cosmology with power-law inflation} 

\author{G.~G.~Luciano}
\email{giuseppegaetano.luciano@udl.cat}
\affiliation{Applied Physics Section of Environmental Science Department, Universitat de Lleida, Av. Jaume
II, 69, 25001 Lleida, Spain}

\date{\today}

\begin{abstract}
We study the inflationary era of the Universe in a modified
cosmological scenario based on the gravity-thermodynamics
conjecture with Barrow entropy instead of the usual Bekenstein-Hawking one. The former arises from the effort to account for quantum gravitational effects on the horizon surface of black holes and, in a broader sense, 
of the Universe. First, we extract modified Friedmann equations from the first law of thermodynamics applied to the apparent horizon of a Friedmann-Robertson-Walker Universe in $(n+1)$-dimensions. Assuming a power-law behavior for the scalar inflaton field, we then investigate how the inflationary dynamics is affected in Barrow cosmological setup.  We find that the inflationary era
may phenomenologically consist of the slow-roll phase, while Barrow entropy is incompatible with kinetic inflation. By demanding observationally consistency of the scalar spectral index and tensor-to-scalar ratio with recent Planck data, we finally constrain Barrow exponent to $\Delta\lesssim10^{-4}$.  
\end{abstract}

 \maketitle

\section{Introduction}
\label{Intro}
The effort to understand the statistical mechanics 
of black holes~\cite{Carlip:2014pma} 
has opened up new scenarios in modern theoretical physics, 
including the study of the AdS/CFT correspondence~\cite{Malda,Aharony:1999ti}
and the investigation of the connection between
gravity and thermodynamics. Beyond their intrinsic interest, 
both these two lines of research might potentially have a
deep impact upon the development of quantum gravity, 
mainly because they are the most successful realizations
of the holographic principle~\cite{tHooft,Sussk}. While the
AdS/CFT correspondence is based on the description of the background 
geometry in terms of anti-de Sitter vacuum solutions, 
the interplay between gravity and thermodynamics 
finds its conceptualization in the so-called \emph{gravity-thermodynamics} conjecture~\cite{Padmanabhan:2003gd,Eling:2006aw,Padmanabhan:2009vy},
which states that Einstein field equations are nothing but the gravitational counterpart of the laws of thermodynamics applied to spacetime~\cite{Jacobson:1995ab}. Besides, in the cosmological context 
such a conjecture allows to extract Friedmann equations by implementing 
the first law of thermodynamics on the apparent horizon of the 
Universe~\cite{Frolov:2002va,Cai:2005ra,Akbar:2006kj,Cai:2006rs}.

In the original formulation the gravity-thermodynamic
conjecture applies Bekenstein-Hawking (BH) area law 
$S_{BH}=A/A_0$ to the Universe apparent 
horizon of surface area $A=4\pi r_{hor}^2$ and radius $r_{hor}$.\footnote{Here and
henceforth we work in natural units $\hslash=c=G=k_B=1$. Accordingly, the Planck area is $A_0=4G=4$.} 
Nevertheless, generalized forms of BH entropy have been discussed
in recent literature, motivated by either nonextensive~\cite{Tsallis,Tsallis2} or quantum gravity~\cite{Barrow} arguments.
To the latter class belongs Barrow entropy, which 
deforms BH area-law to
\be
\label{BE}
S\,=\,\left(\frac{A}{A_0}\right)^{1+\Delta/2}\,,\qquad 0\le\Delta\le1\,,
\ee
where Barrow exponent $\Delta$ 
embeds quantum gravitational corrections. 
In particular, $\Delta=1$ corresponds to the maximal
departure from BH entropy, which is instead recovered for 
$\Delta=0$. Though being proposed for black holes~\cite{Barrow}, 
Eq.~\eqref{BE} is also applied within the cosmological framework, giving rise to modified Friedmann equations that predict
a richer phenomenology comparing to the 
standard one~\cite{Saridakis:2020lrg}. In addition, 
one can rephrase the holographic principle in terms of Barrow entropy, 
obtaining Barrow holographic dark energy (BHDE) (see, for instance
\cite{Saridakis:2020zol,Dabrowski:2020atl,Saridakis:2020cqq,Sheykhi:2021fwh,
Adhikary:2021xym,Luciano,Luciano2,Luciano3} for recent applications). Comparison of the above constructions with
observations sets upper limits on Barrow exponent~\cite{Anagnostopoulos:2020ctz,Leon:2021wyx,Barrow:2020kug,Jusufi:2021fek,LucSar}, 
which slightly deviates from zero, as expected. 

In physical cosmology, inflation is supposed to be 
a crucial era in the evolution of the Universe,
consisting of a very short-lived, but extremely accelerated
expansion phase occurred right after the Big Bang. Originally proposed in~\cite{Inf1,Inf2,Inf3,Inf4}, it has been getting increasing attention
over the years, becoming one of the two pillars of the present
cosmological model along with the late time acceleration~\cite{Perl,Capoz,Salu}.
In spite of this, the origin of inflation 
has not been well understood yet. The most commonly adopted scenario
is that it has been driven by a 
particular form of dark energy represented by a  scalar field with slow rolling assumptions~\cite{LindeCosm}. Alternative models have been recently proposed in~\cite{Alt1,Alt2,Alt3,Alt4,Alt5,Alt6,Alt7,Alt8}. 
The inflationary phase has also been studied in connection with 
holographic dark energy~\cite{InflatHDE1,InflatHDE2,InflatHDE3}, motivated by the plausible role
of the latter as a mechanism responsible for the late time cosmic acceleration.

Starting from the above premises, in this work we study
the evolution and inflation of the Universe
in the context of Barrow entropy-based Cosmology. In this sense, 
our analysis should be regarded as a preliminary attempt
to explore the effects of quantum gravity on the dynamics
of the Universe. In particular, we apply Barrow formula~\eqref{BE} to the entropy associated with the apparent horizon of a $(n+1)$-dimensional 
homogeneous and isotropic (Friedmann Robertson Walker-like) Universe, 
assuming that the matter inside the horizon is 
represented by a scalar field with a potential. In this setting, modified
Friedmann equations are derived from the first law of thermodynamics 
and compared with 
the result of~\cite{SheFr} for the specific case of $n=3$. 
Furthermore, we investigate the early inflationary dynamics of Barrow cosmology with the power-law potential function. 
Contrary to nonextensive (Tsallis-like)
scenario~\cite{Keskin}, where it has been shown that 
inflation may consist of both slow roll- and kinetic-phases, here we find that only the first stage is eligible, the kinetic energy era being 
incompatible  with the allowed values of Barrow exponent $\Delta$. 
After computing the characteristic inflation parameters, 
we infer an upper bound on $\Delta$ in compliance with recent observational constraints on the scalar spectral index and the tensor-to-scalar ratio. We finally comment on the consistency of our results with 
other approaches in literature aimed at exploring inflation driven
by BHDE. 

The remainder of the work is structured as follows:
in the next Section, we derive modified Friedmann equations
from Barrow entropy. Sec.~\ref{IBHDE} is devoted to 
to the study of the inflationary era in BHDE, while conclusions
and outlook are summarized in Sec.~\ref{DaC}.

\section{Modified Friedmann equations in Barrow Cosmology}
\label{MFE}
Let us consider a homogenous and isotropic 
Friedmann-Robertson-Walker (FRW) Universe of spatial curvature $k$. 
We first set notation by following~\cite{Sheykhi:2021fwh} and focusing on $(3+1)$-dimensions. To be as general as possible, the derivation of the modified Friedmann equations in Barrow Cosmology is then performed for the $(n+1)$-dimensional case, with $n\ge3$. 

For a $(3+1)$-dimensional FRW Universe, the line element
can be written as
\be
\label{FRW}
ds^2\,=\,h_{bc}\hspace{0.2mm}dx^{b}dx^{c}+\tilde r^2\left(d\theta^2+\sin^2\theta\,d\phi^2\right),
\ee
where we have denoted the metric
of the $(1+1)$-dimensional subspace by $h_{bc}=\mathrm{diag}[-1,a^2/(1-kr^2)]$. Moreover, $x^b=(t,r)$, $\tilde r=a(t)r$, $a(t)$ is the (time-dependent) scale factor  and $r$ the comoving radius.  

Following~\cite{Akbar2}, the dynamical  
apparent horizon is obtained from the geometric condition 
\be
h^{bc}\partial_b\tilde r \,\partial_c \tilde r=0\,.
\ee 
For the FRW Universe~\eqref{FRW}, explicit calculations yield
\begin{equation}
\label{rtilde}
\tilde r_A=\frac{1}{\sqrt{H^2+{k}/{a^2}}}\,,
\end{equation}
where $H=\dot a(t)/a(t)$ is the Hubble parameter
and the overhead dot indicates derivative respect to the cosmic time $t$.

The apparent horizon has an associated temperature
\be
\label{T}
T=\,\frac{\kappa}{2\pi}=-\frac{1}{2\pi \tilde r_A}\left(1-\frac{\dot{\tilde r}_A}{2H\tilde r_A}\right), 
\ee
where $\kappa$ represents the surface gravity. 
Clearly, for $\dot{\tilde r}_A\le2H\tilde r_A$ we have $T\le0$.  
To avoid meaningless negative temperatures,  one can define 
$T = |\kappa|/2\pi$.  Furthermore, it is possible to assume
that $\dot{\tilde r}_A\ll2H\tilde r_A$ in an  
infinitesimal time interval $dt$,  which amounts
to keeping the apparent horizon radius
fixed. This implies the approximation $T\simeq 1/{2\pi\tilde r_A}$~\cite{Cai:2005ra}. 

We now suppose that
the matter content of the Universe is represented
by a scalar field $\phi$ characterized by a perfect fluid form.
The corresponding Lagrangian is given by
$\mathcal{L}_\phi=X-V(\phi)$, where $X=-\frac{1}{2}h^{\mu\nu}\partial_\mu\phi\partial_\nu\phi$ and $V(\phi)$ are the kinetic
and (spatially homogenous) potential terms, respectively. In turn, 
the stress-energy tensor is
\begin{equation}
\label{SET}
T_{\mu\nu}=(\rho_\phi+p_\phi)u_\mu u_{\nu}+p_\phi\hspace{0.2mm} h_{\mu\nu}\,,
\end{equation} 
where $u_{\mu}$ is the four-velocity of the fluid and
\begin{subequations}
\label{def1}
\begin{equation}
\label{def1bis}
\rho_\phi=\frac{\dot\phi^2}{2}+V(\phi)\,,
\end{equation}
\begin{equation}
\label{def1ter}
p_\phi=\frac{\dot\phi^2}{2}-V(\phi).
\ee
\end{subequations}
represent its energy density and pressure, respectively~\cite{Alt8}. 
In turn, the conservation equation
$\nabla_{\mu}T^{\mu\nu}=0$ gives the continuity equation
\begin{equation}
\label{ce}
\dot \rho_\phi+3H(\rho_\phi+p_\phi)=0\,.
\end{equation}

Combining Eqs.~\eqref{def1} and~\eqref{ce}, we obtain
the dynamics equation of the canonical scalar field as
\be
\label{KG}
\ddot\phi+3H\dot \phi+\partial_\phi V=0\,,
\ee
where the term containing the Hubble constant serves as a kind of friction term resulting from the expansion. 
 
 \subsection{Modified Friedmann equations in $(n+1)$ dimensions}
 \label{MFEndim}
 
The above ingredients provide the basics 
to derive the modified Friedmann equations
in Barrow entropy-based Cosmology. Here, we
extract such equations from the first law of thermodynamics
\be
\label{FLT}
dE\,=\,T\hspace{0.2mm}dS+W\hspace{0.2mm} dV\,,
\ee
applied to the apparent horizon of the FRW Universe in $(n+1)$-dimensions, where 
\be
W=(\rho_\phi-p_\phi)/2\,,
\ee 
is the work density associated
to the Universe expansion and 
\be
\label{engen}
S\,=\,\gamma\left(\frac{A}{A_0^{(n-1)/2}}\right)^{1+\Delta/2}\,,
\ee
is the generalized Barrow entropy. 
We have denoted the $n$-dimensional horizon surface by
$A=n\Omega_n\tilde r_A^{n-1}$, where  
$\Omega_n\,\equiv\,\frac{\pi^{n/2}}{\Gamma(n/2+1)}$
is the angular part of the $n$-dimensional sphere
and $\Gamma$ the Euler's function.
The dimensionless constant $\gamma$ is such that $\gamma\rightarrow1$ for $n=3$, so that  Eq.~\eqref{BE} is restored in this limit. Its explicit expression shall be fixed later. In passing, we mention that 
an alternative derivation of modified Friedmann equations
can be built upon Padmanabhan's paradigm of emergent gravity~\cite{Pad1}, 
which states that the spatial expansion
of our Universe can be understood as the consequence
of the emergence of space with the progress of cosmic time.

Now, by taking into account that the total energy of the Universe inside
the $n$-dimensional volume $V=\Omega_n\tilde r_A^n$
is $E=\rho_\phi V$, we have
\be
\label{Edif}
dE= Vd\rho_\phi + \rho_\phi dV=\Omega_n \tilde r_A^n\hspace{0.2mm} \dot\rho_\phi \hspace{0.5mm}dt \,+\, \rho_\phi\hspace{0.4mm} \Omega_n n\tilde r_A^{n-1}d\tilde r_A\,. 
\ee
This relation can be further manipulated by resorting to the
generalized continuity equation
\begin{equation}
\label{cegen}
\dot \rho_\phi+nH(\rho_\phi+p_\phi)=0\,,
\end{equation}
to give
\be
dE\,=\,-\Omega_n \tilde r_A^n\hspace{0.2mm} n H \left(\rho_\phi+p_\phi\right) dt \,+\, \rho_\phi\hspace{0.4mm} \Omega_n n\tilde r_A^{n-1}d\tilde r_A\,.
\ee

On the other hand, by differentiating the entropy~\eqref{engen}
we get
\begin{eqnarray}
\label{diffS}
dS= \gamma\left(\frac{1}{A_0^{(n-1)/2}}\right)^{1+\Delta/2}n\Omega_n\left(1+\frac{\Delta}{2}\right)\left(n-1\right) \left(n\Omega_n\hspace{0.2mm}\tilde r_A^{n-1}\right)^{\Delta/2}\tilde r_A^{n-2}\hspace{0.2mm}d\tilde r_A\,.
\end{eqnarray}
By plugging Eqs.~\eqref{Edif}-\eqref{diffS} into~\eqref{FLT}, we arrive to
\be
H\left(\rho_\phi+p_\phi\right)dt=\frac{\gamma\left(n-1\right)\left(1+\frac{\Delta}{2}\right)\left(n\Omega \tilde r_A^{n-1}\right)^{\Delta/2}}{2\pi\tilde r_A^3}\left(\frac{1}{A_0^{(n-1)/2}}\right)^{1+\Delta/2}\,dr\,.
\ee
With the further use of the continuity equation~\eqref{cegen}, this becomes
\be
-\frac{2\pi\left(A_0^{(n-1)/2}\right)^{1+\Delta/2}}{\gamma\,n\left(n-1\right)(1+\frac{\Delta}{2})\left(n\Omega\right)^{\Delta/2}}
\,d\rho_\phi\,=\,\tilde r_A^{(n-1)\Delta/2-3}d\tilde r_A\,.
\ee

Integrating both sides, we are led to
\be
\label{40}
\tilde r_A^{(n-1)(1+\Delta/2)-n-1}=\frac{\pi\left[4-\left(n-1\right)\Delta\right]\left(A_0^{(n-1)/2}\right)^{1+\Delta/2}}{\gamma\hspace{0.4mm}n\left(n-1\right)\left(1+\frac{\Delta}{2}\right)\left(n\Omega\right)^{\Delta/2}}\,\rho_\phi\,,
\ee
where the integration constant has been fixed by 
imposing the boundary condition 
$8\pi\rho_{\phi}=\Lambda\simeq0$.
Finally, with the help of the definition~\eqref{rtilde}, we obtain
\be
\label{MFEN}
\left(H^2+\frac{k}{a^2}\right)^{1-(n-1)\Delta/4}\,=\,\frac{8\pi\hspace{0.2mm}G_{eff}^{(n-1)/2}}{3}\hspace{0.2mm}\sigma\hspace{0.2mm}\rho_\phi\,,
\ee
where we have defined
\be
\sigma\equiv\frac{3}{n-2}\,\frac{\left[n+1-\left(n-1\right)\left(1+\frac{\Delta}{2}\right)\right]}{\hspace{0.2mm}n\left(2-\Delta\right)}\,,
\label{sigma}
\ee
and we have set
\be
\label{gamma}
\gamma\,=\,\frac{\pi^{(n-1)\Delta/4}}{2\left(n\Omega_n\right)^{\Delta/2}\,4^{(1+\Delta/2)(1-n)/2}}\,\frac{\left(n-2\right)}{\left(n-1\right)}\left(\frac{2-\Delta}{2+\Delta}\right)^{(3-n)/2}\,.
\ee
Furthermore, we have introduced the effective gravitational constant~\cite{Sheykhi:2021fwh}
\be
G_{eff}\,=\,\frac{A_0}{4}\left(\frac{2-\Delta}{2+\Delta}\right)\left(\frac{A_0}{4\pi}\right)^{\Delta/2}\,.
\ee

Some comments are in order here: first, we notice that for $n=3$, we have 
$\gamma\rightarrow1$, consistently with the discussion below Eq.~\eqref{engen}. The same is true for 
$\sigma$, so that Eq.~\eqref{MFEN} for $n=3$ becomes
\be
\label{mdfe1}
\left(H^2+\frac{k}{a^2}\right)^{1-\Delta/2}\,=\,\frac{8\pi G_{eff}}{3}\hspace{0.2mm}\rho_\phi\,.
\ee
This is nothing but the first modified Friedmann equation
derived in~\cite{Sheykhi:2021fwh} when $\rho_\phi\equiv\rho$ (normal matter).
Furthermore, the limit $\Delta\rightarrow0$ correctly reproduces
the standard Friedmann equation
\be
\label{SFe} 
H^2+\frac{k}{a^2}\,=\,\frac{8\pi}{3}\hspace{0.2mm}\frac{A_0}{4}\hspace{0.2mm}\rho_\phi\,.
\ee

As a final remark, it must be emphasized that, due to the positive
definition of the energy density, Eqs.~\eqref{MFEN} and~\eqref{sigma}
imply the upper bound
\be
n+1-\left(n-1\right)\left(1+\frac{\Delta}{2}\right)>0 \,\,\,\, \Longrightarrow\,\,\,\, \Delta<\frac{4}{n-1}\,,
\ee
which is obviously satisfied for any allowed value of $n$. 

Now, from the time derivative of Eq.~\eqref{mdfe1}, one can
easily obtain the second modified Friedmann equation as follows
\be
2H\left[1-\left(n-1\right)\frac{\Delta}{4}\right]\left(H^2+\frac{k}{a^2}\right)^{-(n-1)\Delta/4}\left(\frac{\ddot a}{a}-H^2-\frac{k}{a^2}\right)=\frac{8\pi G_{eff}^{(n-1)/2}}{3}\sigma\dot\rho_\phi\,.
\ee
By use of the continuity equation~\eqref{cegen}, this gives
\be
\left[1-\left(n-1\right)\frac{\Delta}{4}\right]\left(H^2+\frac{k}{a^2}\right)^{-(n-1)\Delta/4}\left(\frac{\ddot a}{a}-H^2-\frac{k}{a^2}\right)=-\frac{4\pi G_{eff}^{(n-1)/2}}{3}\,\sigma n \left(\rho_\phi+p_\phi\right).
\ee
Replacing $\rho_\phi$ by the first Friedmann equation~\eqref{MFEN},
we find after some simplification
\be
\left[4-\left(n-1\right)\Delta\right]\frac{\ddot a}{a}\left(H^2+\frac{k}{a^2}\right)^{-(n-1)\Delta/4}+\left[2n-4+\Delta\left(n-1\right)\right]\left(H^2+\frac{k}{a^2}\right)^{1-(n-1)\Delta/4}=-\frac{16\pi G_{eff}^{(n-1)/2}}{3}\,\sigma\hspace{0.2mm} n\hspace{0.2mm} p_\phi\,.
\ee
This is the second modified Friedmann equation in Barrow Cosmology.
Again, one can check that $n=3$ gives back the
result of~\cite{Sheykhi:2021fwh}
\be
\label{mdfe2}
\left(2-\Delta\right)\frac{\ddot a}{a}\left(H^2+\frac{k}{a^2}\right)^{-\Delta/2}+
\left(1+\Delta\right)\left(H^2+\frac{k}{a^2}\right)^{1-\Delta/2}\,=\,-8\pi G_{eff}\,p_\phi\,.
\ee
The further limit $\Delta\rightarrow0$ reproduces the standard second Friedmann equation, here rewritten as
\be
\label{sfem}
\dot H+H^2=-\frac{4\pi}{3}\left(\rho_\phi+3p_\phi\right), 
\ee
where we have used the relation 
\be
\dot H=\frac{\ddot a}{a}-H^2\,.
\ee

\section{Inflation in Barrow Cosmology}
\label{IBHDE}

Let us now move onto the study of the inflationary 
era of the Universe. Within the scalar theory framework considered
above, the characteristic quantities to compute are the 
inflation slow-roll parameters, which are defined by
\begin{eqnarray}
\label{slow1}
\epsilon &=& - \frac{\dot H}{H^2}\,,\\[2mm]
\eta&=&-\frac{\ddot H}{2H\dot H}\,.
\label{slow2}
\end{eqnarray}

Slow-roll conditions assert that both these
two parameters take very small values during inflation, i.e.
$\epsilon,\eta\ll1$. 
In the slow-roll theoretical framework, 
the only requirement $\epsilon\ll1$ is actually needed 
to ensure the existence of an early inflationary era, 
Then, by imposing $\dot\phi^2,\ddot \phi\ll1$ on the equation of motion of the theory, the first Friedmann equation~\eqref{MFEN}
under the slow-roll assumptions becomes
\be
\label{novel1}
H^2\simeq \left[\frac{8\pi G_{eff}}{3}\, V(\phi)\right]^{2/(2-\Delta)},
\ee
where we have focused on the case $n=3$
and we have resorted to Eq.~\eqref{def1bis}. 

On the other hand, from the second Friedmann equation~\eqref{mdfe2}
we get
\be
\label{novel2}
\dot H\simeq\frac{3\dot\phi^2}{2\left(\Delta-2\right)}\left(\frac{8\pi G_{eff}}{3}\right)^{2/(2-\Delta)}V(\phi)^{\Delta/(2-\Delta)}\,.
\ee 
Combining Eqs.~\eqref{novel1} and~\eqref{novel2}, 
the slow-roll parameters~\eqref{slow1} and~\eqref{slow2}
take the form
\begin{eqnarray}
\label{eps36}
\epsilon&\simeq&\frac{3\dot\phi^2}{2\left(2-\Delta\right)}V(\phi)^{-1}\,,\\[2mm]
\eta&\simeq& -\left(\frac{8\pi G_{eff}}{3}V(\phi)\right)^{1/(\Delta-2)}
\left[\frac{\ddot \phi}{\dot\phi}+\frac{\dot\phi\,\Delta}{4-2\Delta}\frac{\partial_\phi V(\phi)}{V(\phi)}\right].
\label{eta36}
\end{eqnarray}
Let us now remark that the above parameters should be
computed at horizon crossing, where the fluctuations of the inflation field
freeze~\cite{Keskin}. 

The scalar spectral index of the primordial curvature
perturbations and the tensor-to-scalar ratio are defined by
\begin{eqnarray}
\label{ns}
n_s&\simeq& 1-6\epsilon+2\eta\,,\\[2mm]
r&\simeq&16\epsilon\,,
\label{ttsr}
\end{eqnarray}
respectively, which also need to be evaluated at the horizon crossing.
For later convenience, it is useful to introduce the
e-folding time
\be
\label{N}
N=\int_{t_i}^{t_f} H(t) dt\,,
\ee
where $t_i$ $(t_f)$ represents the initial (final) time
of the inflationary era. Consistently with the above discussion, 
we consider $t_i=t_c$ as the horizon crossing time, 
so that Eq.~\eqref{N} can be rewritten as $N=\int_{\phi_c}^{\phi_f} H{\dot\phi}^{-1} \,d\phi$, where we have used the notation $\phi_c\equiv \phi (t_c)$ and 
$\phi_f\equiv \phi (t_f)$. 

\subsection{Slow-roll inflation with power-law potential}
We now examine inflation from the dynamical point of view. 
Toward this end, we assume a power-law behavior for the scalar
potential $V(\phi)$ in the form
\be
\label{potential}
V(\phi)\simeq\phi^{m}\,,
\ee
where $m>0$ is the power-term.
The latest observational data prefer
models with $m\sim\mathcal{O}(1)$ or $m\sim\mathcal{O}(10^{-1})$, 
while $m\ge2$ is disfavored in the minimally coupled scalar field. 
Henceforth, we shall focus on such phenomenologically allowed values
of $m$.
We also remark that power law inflation is a very 
useful model to assess approximation schemes for the computation
of scalar power spectra, since its spectrum is exactly solvable\footnote{More generally, one can assume $V(\phi)=V_0\phi^{m}$, where $V_0$ is a positive constant with dimensions of $[E]^{4-m}$. However, since observational indices are shown to be independent of this quantity, we can set $V_0=1$ in suitable units without loss of generality.}.

In order to extract analytical
solutions of the inflationary observable indices, 
we express $\dot\phi$ and $\ddot\phi$ in terms
of the scalar field by using the slow-roll conditions.
In this regard, let us observe that the evolution equation~\eqref{KG}
can be rewritten as
\be
\dot\phi\simeq-\frac{1}{3H}\hspace{0.2mm}\partial_\phi V\,. 
\ee
By plugging into~\eqref{novel1}, we get 
\be
\label{phidot}
\dot\phi=-\frac{m}{3}\hspace{0.2mm}\left(\frac{8\pi G_{eff}}{3}\right)^{1/(\Delta-2)}\,\phi^{\left[\left(2-\Delta\right)\left(m-1\right)-m\right]/\left(2-\Delta\right)}\,.
\ee

We can now derive the expression of $\phi_f$ by
noticing that inflation is supposed to end when $\epsilon(\phi_f)\sim1$.
By inverting Eq.~\eqref{eps36}, we are led to
\be
\phi_f=\left[\frac{6\left(2-\Delta\right)}{m^2}\hspace{0.2mm}\left(\frac{8\pi G_{eff}}{3}\right)^{2/(2-\Delta)}
\right]^{(2-\Delta)/[\Delta(2-m)-4]}\,.
\ee 

Similarly, insertion of Eqs.~\eqref{novel1} and~\eqref{phidot}
into~\eqref{N} allows to infer the following expression for the scalar
field at horizon crossing
\be
\label{phic}
\phi_c=\Bigg\{ \frac{m}{3\left(2-\Delta\right)}\left(\frac{8\pi G_{eff}}{3}\right)^{2/(\Delta-2)}\left\{\frac{m}{2}+N\left[4+\Delta\left(m-2\right)\right]
\right\}\Bigg\}^{(2-\Delta)/\left[4+\Delta(m-2)\right]}\,.
\ee

The scalar spectral index~\eqref{ns} and the tensor-to-scalar ratio~\eqref{ttsr} 
can be cast in terms of the power-term $m$ and the e-folding time $N$
as
\begin{eqnarray}
\label{nsbis}
n_s&\simeq&1-\frac{2\left[4+\Delta(m-2)+m\right]}{m\left\{1+\frac{2N}{m}\left[4+\Delta(m-2)\right]\right\}}\,,\\[2mm]
r&\simeq&\frac{16}{1+\frac{2N}{m}\left[4+\Delta(m-2)\right]}\,.
\label{rbis}
\end{eqnarray}
Remarkably, we see that the slow-roll indices only depend
on the power-term $m$ and Barrow parameter $\Delta$. 
A similar result has been
exhibited in the context of Tsallis deformation of entropy-area law~\cite{Keskin}.

\begin{figure}[t]
\centering
\includegraphics[width=8.8cm]{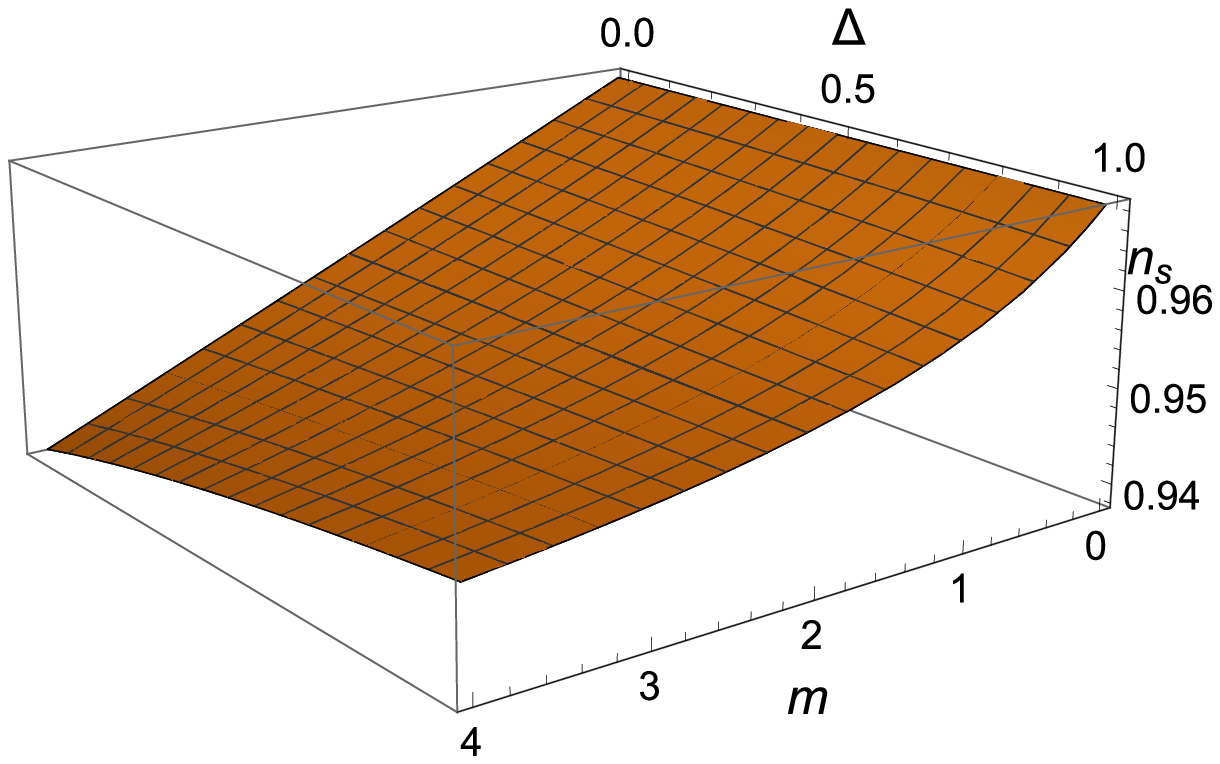}
\includegraphics[width=8.9cm]{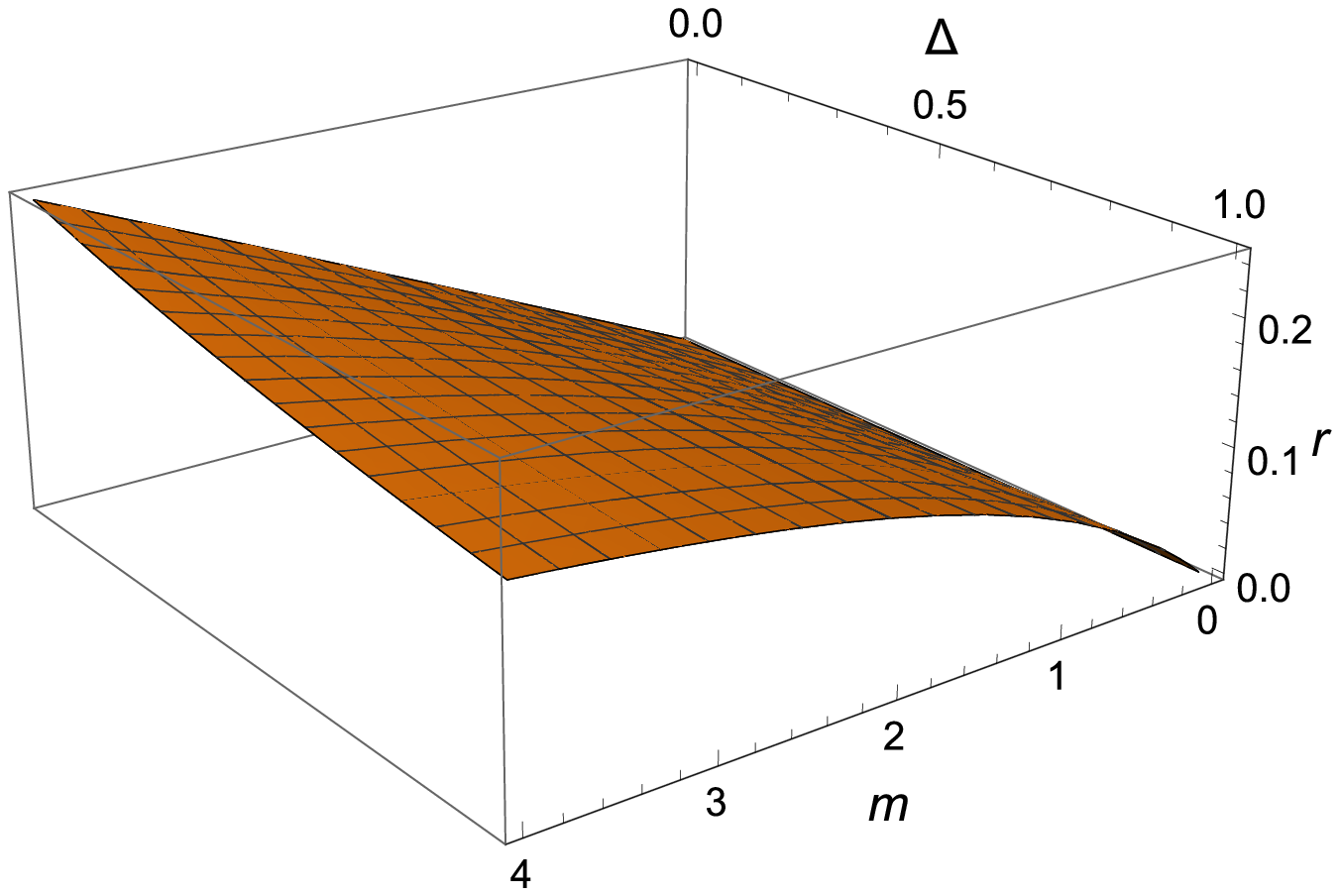}\caption{Plot of $n_s$ (left panel) and $r$ (right panel) versus the power-term $m$ and Barrow parameter $\Delta$ for $N\sim 30$.}
\label{Fig1}
\end{figure}

In order to constrain Barrow exponent $\Delta$, 
let us require consistency of Eqs.~\eqref{nsbis} and~\eqref{rbis}
with observations. Specifically, we consider 
Planck 2018 measurements, which set the following bounds on $n_s$
and $r$~\cite{ConPlanck}
\begin{eqnarray}
\label{nsb}
n_s&=& 0.9649 \pm 0.0042 \,\,\,\, (68\%\,\, \mathrm{CL}) \,\,\,\, \mathrm{from\,\,Planck\,\, TT,TE,EE+lowE+lensing}\,,  \\[2mm]
r&<&0.064 \hspace{1.8cm} (95\%\,\, \mathrm{CL}) \,\,\,\, \mathrm{from\,\, Planck\,\, TT,TE,EE+lensing+lowEB}\,.
\label{rb}
\end{eqnarray}
For a sufficiently long inflation ($N\sim 30$ to $40$ 
e-folds) and $m\sim\mathcal{O}(1)$, 
we obtain $\Delta\lesssim10^{-4}$ (see also Fig.~\ref{Fig1}),  
which improves the constraints $\Delta=0.0094^{+0.094}_{-0.101}$ and
$\Delta\lesssim0.08$ obtained from Supernovae (SNIa) Pantheon sample~\cite{Anagnostopoulos:2020ctz,Leon:2021wyx} and
baryogenesis~\cite{LucSar} measurements, and is consistent with the most stringent bound $\Delta\lesssim1.4\times10^{-4}$ found so far through Big Bang Nucleosynthesis~\cite{Barrow:2020kug}. We also observe that this bound is not improved by considering smaller power-terms $m\sim\mathcal{O}(10^{-1})$.

\subsection{Kinetic inflation with power-law potential}
Above we have argued that the
slow-roll inflation terminates
when $\epsilon\sim 1$. Two scenarios can then occur:
either the scalar field oscillates to the minimal value of the potential,  
leading the Universe into a decelerated expansion phase, 
or the inflation goes on but with different features. Here, we shall examine
whether the latter possibility is allowed within Barrow entropy-based Cosmology. 
In particular, a crucial assumption of the slow-roll inflation is that
the kinetic energy of the scalar field can be neglected. 
However, if the volume of the Universe is large enough before
the field starts to oscillate, then a kinetic term might arise 
and drive a transition from a vacuum state to quintessence.
We assume the kinetic contribute
in the form
\be
\label{kinetic}
\dot\phi^2=m\hspace{0.2mm} V(\phi)\,.
\ee
The above expression
can be actually deduced from the dynamics relation~\eqref{KG} and the modified Friedmann equations~\eqref{mdfe1}
and~\eqref{mdfe2}, here rewritten for convenience as
\begin{eqnarray}
H^2&=&\left(\frac{8\pi G_{eff}}{3}\right)^{2/(2-\Delta)}\left[\frac{\dot\phi^2}{2}+V(\phi)\right]^{2/(2-\Delta)}\,, \\[2mm]
\dot H&=&\frac{3}{\Delta-2}\hspace{0.2mm}\dot\phi^2\left(\frac{8\pi G_{eff}}{3}\right)^{2/(2-\Delta)}\left(\frac{\dot\phi^2}{2}+V\right)^{\Delta/(2-\Delta)}.
\end{eqnarray}
These equations also allow us to express the slow-roll parameters
as
\begin{eqnarray}
\label{eps2}
\epsilon&=&\frac{6m}{\left(2-\Delta\right)\left(m+2\right)}\,,\\[2mm]
\eta&=&\frac{m^{3/2}}{(\Delta-2)}\,
\left(\frac{8\pi G_{eff}}{3}\right)^{1/(\Delta-2)}
\left(\frac{m+2}{2}\right)^{1/(\Delta-2)}
\hspace{0.2mm}
\phi^{\left[\Delta(2-m)-4\right]/\left[2(2-\Delta)\right]}\,.
\label{eta2}
\end{eqnarray}

\begin{figure}[t]
\centering
\includegraphics[width=8.8cm]{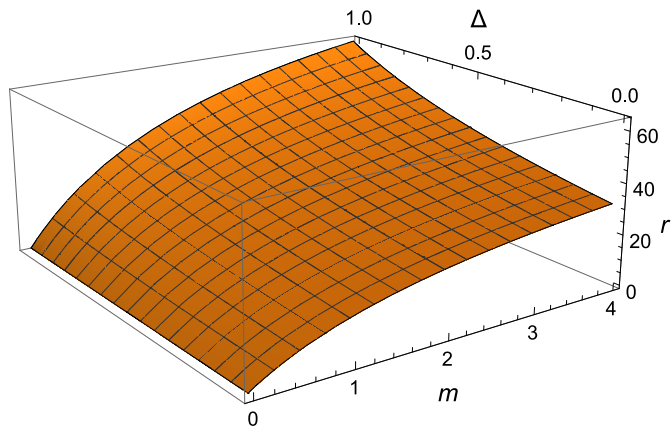}
\caption{Plot of $r$ versus the power-term $m$ and Barrow parameter $\Delta$.}
\label{Fig3}
\end{figure}

Now, the end of the kinetic inflation is 
set by the condition  $\eta(\phi_f)\simeq1$~\cite{Keskin}, 
which gives from the definition~\eqref{N}
\be
\phi_c=\left\{
\left(\frac{8\pi G_{eff}}{3}\right)^{1/(\Delta-2)}
\left(\frac{m+2}{2}\right)^{1/(\Delta-2)}\frac{m^{1/2}}{2(\Delta-2)}
\left[
2m+
N\left[4+\Delta\left(m-2\right)\right]
\right]
\right\}
^{2(2-\Delta)/[4+\Delta(m-2)]}\,.
\ee
From Eqs.~\eqref{eps2} and~\eqref{eta2}, we then get (see also Fig.~\ref{Fig3})
\begin{eqnarray}
\nonumber
n_s&=&\Bigg\{1
+
4m\,
\Bigg\{\frac{9}{\left(m+2\right)\left(\Delta-2\right)}\\[2mm]
&&+\,
\frac{\left\{\frac{m^{1/2}}{2(\Delta-2)}\left(\frac{m+2}{2}\right)^{1/(\Delta-2)}\left(\frac{8\pi G_{eff}}{3}\right)^{1/(\Delta-2)}\left[2m+N\left[4+\Delta\left(m-2\right)\right]\right]\right\}^{n\Delta/[2(\Delta-2)]}}
{2m+N\left[4+\Delta\left(m-2\right)\right]}
\Bigg\}
\Bigg\}^{2(2-\Delta)/[4+\Delta(m-2)]},
\\[2mm]
r&=&\frac{96m}{\left(m+2\right)\left(2-\Delta\right)}\,.
\label{rbisbis}
\end{eqnarray}

Unlike the previous scenario, we now find that
observationally consistency is obtained, provided
that $\Delta$ assumes largely negative values. This 
occurs for both $m\sim\mathcal{O}(1)$ and $m\sim\mathcal{O}(10^{-1})$,
as it can be easily seen from Eq.~\eqref{rbisbis}.
However, such a condition is at odds with the assumption~\eqref{BE}, 
implying that a kinetic inflation could not be explained within 
Barrow's framework. This is a remarkable difference with
the case of inflation based on Tsallis entropy~\cite{Keskin}, 
which allows for kinetic phase too. Specifically, in that case
the kinetic inflation is associated to a regime
of decreasing horizon entropy and ensuing clumping
of fluctuations in particular regions of spacetime.

\section{Discussion and Conclusions}
\label{DaC}
Inspired by Covid-19 fractal structure, 
the modified entropy-area law~\eqref{BE} has
been proposed to take into account quantum gravitational effects
on the black hole horizon surface~\cite{Barrow}. 
In the lines of the gravity-thermodynamic conjecture, this paradigm
has been applied to the Universe horizon too, the ensuing
framework being known as Barrow Cosmology.
Within this framework, we have studied the 
evolution of FRW Universe, assuming
the matter content to be represented by a homogeneous
scalar field in the form of a perfect fluid. As a first step, by using
the first law of thermodynamics applied to the horizon of the FRW Universe, 
we have derived modified ($\Delta$-dependent) Friedmann equations.
The obtained result has been used to analyze the inflationary era. 
Toward this end, we have supposed a power-law behavior 
for the scalar inflaton field. We have found that inflation in Barrow Cosmology can consist of the slow-roll phase only,
the kinetic inflation being incompatible with the allowed values of Barrow deformation parameter.
We have finally constrained Barrow exponent to $\Delta\lesssim10^{-4}$ by demanding consistency of the scalar spectral index and tensor-to-scalar ratio with recent observational Planck data.

Other aspects deserve further analysis. Besides the background
and inflationary evolution, it would be interesting to study
the growth rate of matter density perturbations and structure formation. This is an important testing ground to discriminate
among existing modified cosmological models. 
Preliminary investigation in this direction has been proposed in~\cite{Perturb} in the context of both Tsallis and Barrow entropies, showing
that the entropic deformation parameter significantly  
influences the growth of perturbations. Moreover, 
one can attempt to extend the present considerations
to Cosmology based on Kaniadakis entropy~\cite{Kan1}, 
which is a self-consistent relativistic generalization of Boltzmann-Gibbs entropy with non-trivial cosmological implications~\cite{LucRev}. In this way, a relationship between Barrow and Kaniadakis
formalisms can be established.  
Finally, since our models is an effort to include
quantum gravity corrections in the analysis of inflation, 
it is essential to examine the obtained results
in connection with predictions from more fundamental
theories of quantum gravity~\cite{Addazi}. Work along these and other
directions is under active consideration and will be
presented elsewhere.

\bigskip 

\noindent \textbf {Acknowledgements}
The author acknowledges the Spanish ``Ministerio de Universidades'' 
for the awarded Maria Zambrano fellowship and funding received
from the European Union - NextGenerationEU. 
He is also grateful for participation in the COST
Action CA18108  ``Quantum Gravity Phenomenology in the Multimessenger Approach''.

\end{document}